%% file: DAC2025_draft.tex
\documentclass[conference]{IEEEtran}

\def\BibTeX{{\rm B\kern-.05em{\sc i\kern-.025em b}\kern-.08em
    T\kern-.1667em\lower.7ex\hbox{E}\kern-.125emX}}

\usepackage{booktabs} 

\usepackage{amsmath,amsfonts}
\usepackage{algorithm}
\usepackage{algorithmicx}  
\usepackage{algpseudocode}
\usepackage{multirow}

\usepackage{array}
\usepackage[caption=false,font=normalsize,labelfont=sf,textfont=sf]{subfig}
\usepackage{textcomp}
\usepackage{stfloats}
\usepackage{url}
\usepackage{verbatim}
\usepackage{graphicx}
\usepackage{cite}
\usepackage{xcolor}  

\begin{document}

\title{A Tensor-Train Decomposition based Compression of LLMs on Group Vector Systolic Accelerator}

\author{Sixiao~Huang, 
        Tintin~Wang, 
        Ang~Li, 
        Ao~Shen, 
        Kai~Li, 
        Keyao~Jiang, 
        Mingqiang~Huang 
		and~Hao~Yu~\IEEEmembership{Senior~Member,~IEEE}

\thanks{This work was supported by STI 2030, Major Projects (2022ZD0210600), National Natural Science Foundation of China (NSFC) (Key Program Grant No. 62034007), the Key-Area Research and Development Program of Guangdong Province (Grant No. 2019B010142001) and Shenzhen Science and Technology Program (Grant No. KQTD20200820113051096 and JCYJ20200109115210307). (Corresponding author: Hao Yu).}
\thanks{Mingqiang Huang is with Shenzhen Institute of Advanced Technology, Chinese Academy of Sciences, Shenzhen 518055, China.}
\thanks{Sixiao Huang, Tintin Wang, Ang Li, Ao Shen, Kai Li, Keyao Jiang and Hao Yu are with School of Microelectronics, Southern University of Science and Technology, Shenzhen 518055, China. (yuh3@sustech.edu.cn)}}

\maketitle

\begin{abstract}
Large language models (LLMs) are both storage-intensive and computation-intensive, posing significant challenges when deployed on resource-constrained hardware. As linear layers in LLMs are mainly resource consuming parts, this paper develops a tensor-train decomposition (TTD) for LLMs with a further hardware implementation on FPGA. TTD compression is applied to the linear layers in ChatGLM3-6B and LLaMA2-7B models with compression ratios (CRs) for the whole network 1.94$\times$ and 1.60$\times$, respectively. The compressed LLMs are further implemented on FPGA hardware within a highly efficient group vector systolic array (GVSA) architecture, which has DSP-shared parallel vector PEs for TTD inference, as well as optimized data communication in the accelerator. Experimental results show that the corresponding TTD based LLM accelerator implemented on FPGA achieves 1.45$\times$ and 1.57$\times$ reduction in first token delay for ChatGLM3-6B and LLaMA2-7B models, respectively. 

\end{abstract}


\section{Introduction}

Artificial intelligence has already demonstrated extraordinary accuracy across numerous tasks, profoundly impacting people's lives\cite{russakovsky2015imagenet, long2015fully, hinton2012deep, vaswani2017attention}. However, deploying deep neural networks (DNNs) on edge devices is very challenging due to limited storage, computing resources, and power consumption. Previous work\cite{huang2022high, chen2016eyeriss, huang2022high3d, yang2024lamps} has focused on accelerating convolutional neural networks (CNNs), primarily optimizing convolutional layers inference, while linear layers are converted to convolutional computations. However, due to the low ratio of operations to weight number of linear layers, the memory access limits for performance of accelerators, as described by the "Roofline" model\cite{zhang2015optimizing}. Transformers and LLMs are developing rapidly, most of whose computation are linear layers. Therefore, solving the issue of large memory access in linear layers is crucial for deploying LLMs on edge devices with efficient performance.

\begin{figure}[t]
  \centering
  \includegraphics[width=\linewidth]{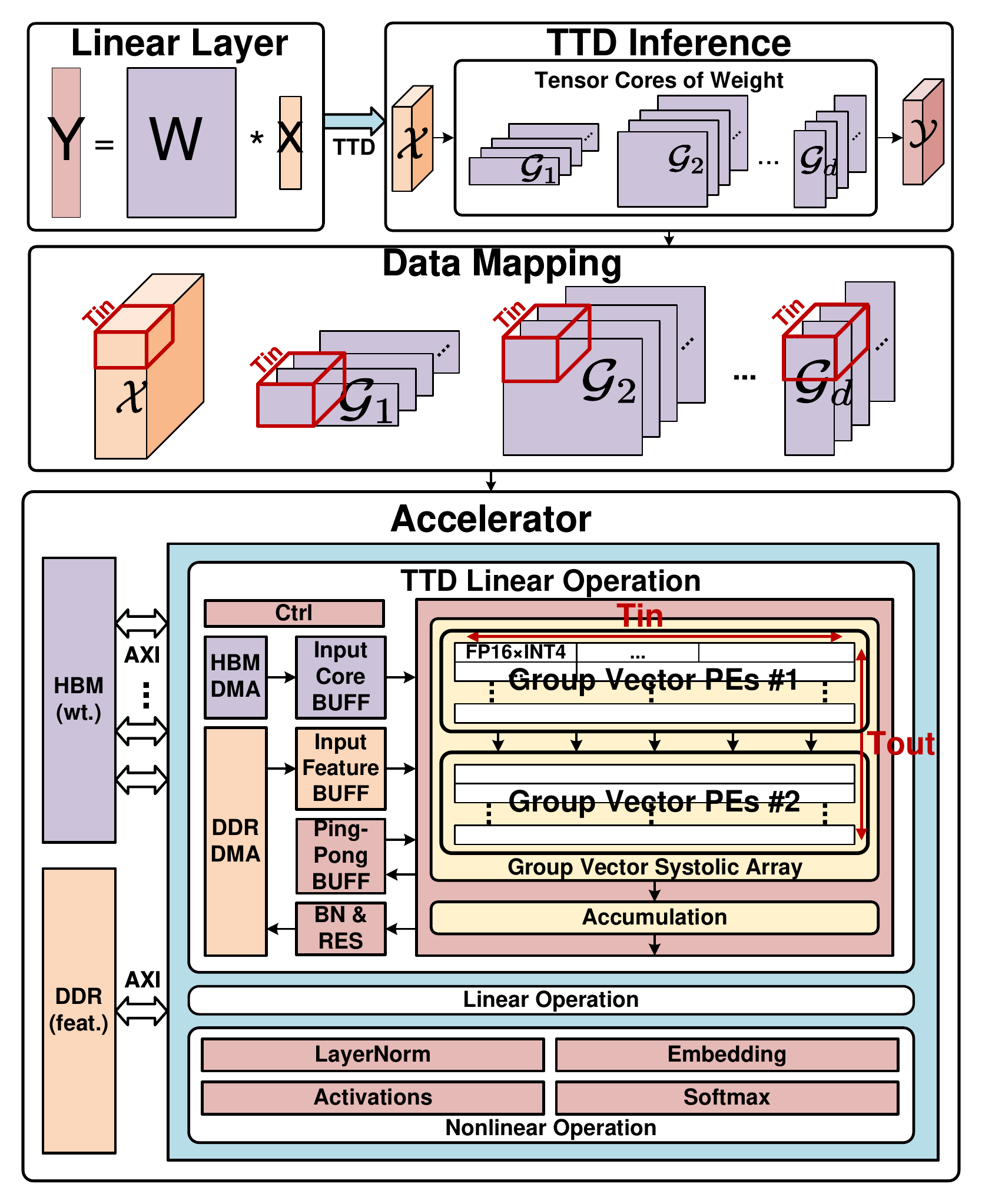}
  \caption{LLMs with TTD compressed linear layers mapping on FPGA implemented group vector systolic accelerator.}
  \label{fig_总体架构}
  \vspace{-0.4cm}
\end{figure}

There are two primary approaches for compressing DNNs to decrease the memory access. The first approach is reducing the bit width of data, known as quantization\cite{chang2021mix, guan2024aptq}. This method converts the neural network weights into integer numbers, which consume fewer bits than the full-precision floating-point values used during network training. This not only reduces the size of the weights but also makes network computations more hardware-friendly, as the integer calculations require significantly fewer hardware resources and power compared to floating-point calculations. Previous work\cite{he2023agile} deployed NAS optimized multi-precision CNN models on FPGA and achieved impressive results. The second approach focuses on reducing the number of weights through techniques such as sparse pruning\cite{liang2020omni} and low-rank compression. These methods decrease the number of weights by changing the network's architecture. Among these compression methods, low-rank compression stands out due to its high compression ratio (CR), especially the tensor-train decomposition (TTD) method\cite{oseledets2011tensor}. Experiments have shown that applying TTD to the linear layers of VGG16 on the ImageNet dataset achieved a CR of 50,000 $\times$\cite{novikov2015tensorizing}. Previous works\cite{man2023ranksearch, cheng2019deepeye, wei2024fmtt} show that TTD performs effectively on RNNs and 3D point cloud models. Despite TTD's remarkable compression capabilities, it changes the computational structure of linear layers, necessitating a redesign of hardware to accelerate efficiently.

In this paper, we propose an FPGA-based accelerator to accelerate TTD compressed LLMs which is shown in Fig.~\ref{fig_总体架构}. First, TTD compression is applied to linear layers in ChatGLM3-6B and LLaMA2-7B with whole-network CRs of 1.94$\times$ and 1.60$\times$, respectively. Second, we implement the TTD inference of the linear layer on hardware. Not only do we need to reduce the redundant computation and memory access in the original TTD inference, but we also need to perform the reordering operation on the intermediate data during the TTD. To achieve optimize hardware efficiency, the accelerator uses a group vector systolic array (GVSA) to perform matrix computations during TTD inference. The DSP-shared parallel vector processing elements (PEs) execute FP16 and INT4 multiply-accumulate (MAC) operations. Through operator fusion, we implement three operations on the accelerator: TTD linear operation, linear operation, and nonlinear operation. By configuring and chaining these operations together, edge deployment of TTD compressed LLMs can be achieved. Finally, we implement the accelerator design on the AMD Alveo V80 FPGA to deploy ChatGLM3-6B and LLaMA2-7B after TTD compression. Compared to the baseline without the TTD inference, the accelerator achieves 1.45$\times$ and 1.57$\times$ reduction in first token delay for ChatGLM3-6B and LLaMA2-7B, respectively. 

\section{TTD Compressed Linear Layer}

\subsection{Linear Layers in LLMs}
\begin{figure}[t]
\vspace{-0.4cm}
  \centering
  \includegraphics[width=0.95\linewidth]{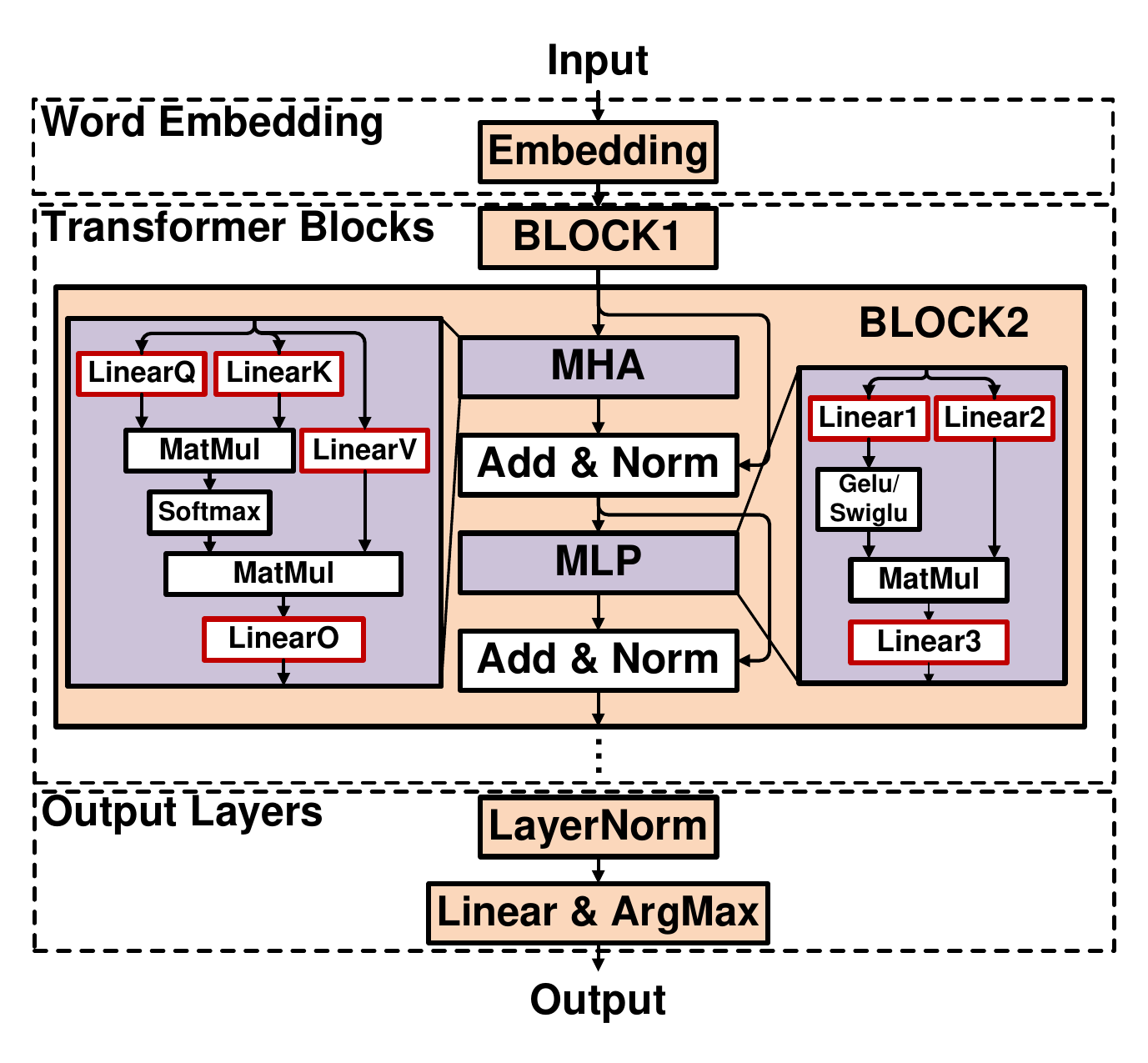}
  \caption{Architecture of ChatGLM3-6B and LLaMA2-7B.}
  \label{fig_FC}
  \vspace{-0.4cm}
\end{figure}

Fig.~\ref{fig_FC} shows the architecture of ChatGLM3-6B\cite{du2021glm} and LLaMA2-7B\cite{touvron2023llama}, which consists of word embedding, transformer blocks, and output layers. The transformer blocks are composed of multiple repeated blocks in sequence. A single block is made up of a multi-head attention (MHA) and  a multilayer perceptron (MLP). The MHA consists of linear layers, matrix multiplication, and softmax, while the MLP is composed of linear layers and activation functions. It is evident that most of the computations in LLMs are linear layer operations. Therefore, to deploy LLMs on edge devices efficiently, it is essential to compress the linear layers to reduce the number of network weights. 

\subsection{Tensorization}
The weights of the linear layers described in the previous section are all two-dimensional matrices, whereas the TTD algorithm is applied to higher-dimensional tensor (dimension greater than 2). Therefore, to apply TTD to compress the weight matrix of the linear layer, it first needs to be tensorized. Tensorization is the process of converting low-dimension data (vector or matrix) into high-dimension data (tensor). For example, there is a matrix $T \in \mathbb{R}^{M_1 \times M_2}$, and its tensorized form is $\mathcal{T} \in \mathbb{R}^{v_1 \times v_2 \times ... \times v_d}$ which is a $d$-dimensional tensor. The number of elements remains the same before and after tensorization. So $M_1M_2= {\textstyle \prod_{i=1}^{d}v_i} $. Through this process, the weight matrix of the linear layer can be transformed into a tensor of arbitrary dimensions, facilitating the subsequent application of TTD.

\subsection{Tensor-Train Decomposition}
\begin{figure}[h]
    \vspace{-0.4cm}
  \centering
  \includegraphics[width=\linewidth]{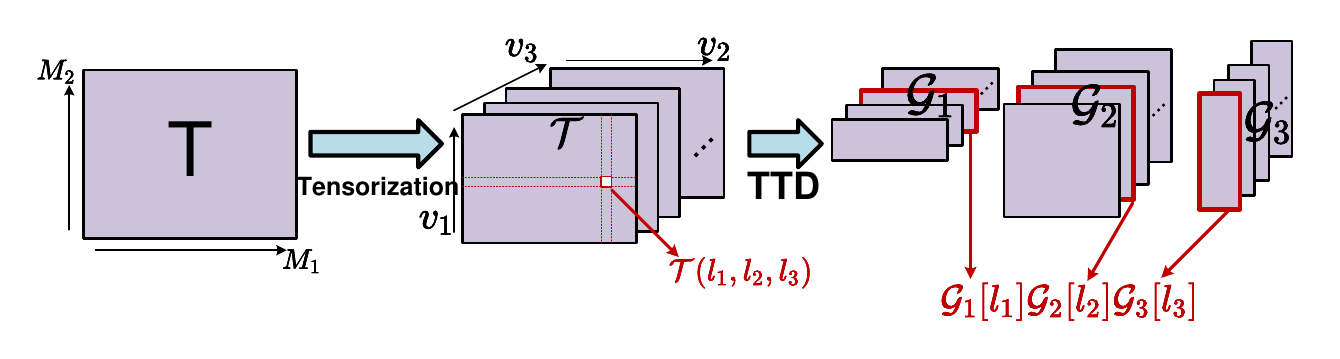}
  \caption{Basic principles of tensorization and TTD (Example with $d=3$).}
  \label{fig_TTD原理}
  \vspace{-0.4cm}
\end{figure}

TTD\cite{oseledets2011tensor} has been proved for compressing multi-dimensional tensors into a more compact form. This method decomposes a high-dimensional tensor into a sequence of lower-dimensional tensors, known as cores, which are multiplied together in a specific order. Specifically, for a $d$-dimensional tensor $\mathcal{T} \in \mathbb{R}^{v_1 \times v_2 \times ... \times v_d}$, it can be decomposed as follows:
\begin{equation}
\mathcal{T}(l_1,...,ld)=\mathcal{G}_1[l_1]\mathcal{G}_2[l_2]...\mathcal{G}_d[l_d]
\end{equation}
where $\mathcal{G}_k[l_k] \in \mathbb{R}^{r_{k-1} \times r_k}$ is the $l_k$ slice of the $k^{th}$ core $\mathcal{G}_k$ and $l_k \in [1,v_k]$ and $r_k$ is the rank. To satisfy the shape of the tensor, $r_0$ and $r_d$ are set as 1. And the CR of TTD is: 
\begin{equation}
CR = \frac{\textstyle \prod_{k=1}^{d}v_k}{\textstyle \sum_{k=1}^{d}v_kr_{k-1}r_k}  
\end{equation}
where the ranks of cores are always small, so the compression effect will be very obvious. The basic principle of tensorization and TTD is shown in Fig.~\ref{fig_TTD原理}.

\subsection{TTD Compression and Inference}
We can perform a TTD compressed linear layers for inference as shown in Algorithm~\ref{alg_TTD算法框} below. The linear layer can be simply expressed as $Y=Wx$ where $Y \in \mathbb{R}^M$, $W \in \mathbb{R}^{M \times N}$ and $x \in \mathbb{R}^N$. First, we need to perform tensorization operations on the input and weight matrices to tensor $\mathcal{W} \in \mathbb{R}^{(m_1 \times n_1)  \times ... \times (m_d \times n_d)}$ and $\mathcal{X} \in \mathbb{R}^{n_1  \times ... \times n_d}$ where $M={\textstyle \prod_{k=1}^{d}} m_{k}$ and $N={\textstyle \prod_{k=1}^{d}} n_{k}$. Note that here,  $d$  corresponds to the number of cores $\mathcal{G}$ generated by TTD later. Additionally, we need to determine the rank of the cores for TTD, which affects the CR of TTD and the similarity between the reconstructed matrix from the cores and the original weight matrix. A higher rank results in a lower CR and higher similarity and it has a smaller impact on network accuracy. The TTD inference of linear layers can be expressed as:
\begin{equation}
\mathcal{Y}(j_1,...,j_d)=\sum_{i_1...id}^{}\mathcal{G}_1[i_1,j_1]\mathcal{G}_2[i_2,j_2]...\mathcal{G}_d[i_d,j_d]\mathcal{X}(i_1,...,i_d)
\label{eq_TTD}
\end{equation}
where the output $\mathcal{Y} \in \mathbb{R}^{m_1 \times m_2 \times ... \times m_d}$ is the high-dimension tensor format of $Y$ after TTD. And $\mathcal{G}_k \in \mathbb{R}^{m_k \times r_{k-1} \times r_k \times n_k}$ is $k^{th}$ core of weight tensor $\mathcal{W}$.

By compressing the weight matrix of the linear layer using TTD, we obtain the corresponding cores. The TTD inference of the linear layer can be performed as described in Equation~(\ref{eq_TTD}). This involves multiplying the tensorized input by the cores of weights of TTD compressed linear layer, and then summing over all input dimensions $i_k$ to obtain the final tensorized output.

\begin{algorithm}[t]
\caption{TTD compression of linear layers}
\label{alg_TTD算法框}
\begin{algorithmic}[1]

\Require
The input and weight of the linear layer, $x, W$; The dimension number of TTD, $d$; The rank of TTD, $r$.

\Ensure 
The input and cores of TTD compressed linear layer, $\mathcal{X}, \mathcal{G}_1, ... \mathcal{G}_d$.

\State $n_1, n_2, ..., n_d = Factorization(N, d)$
\State $m_1, m_2, ..., m_d = Factorization(M, d)$
\State $\mathcal{X} = Tensorization(x, (n_1, ..., n_d))$
\State $\mathcal{W} = Tensorization(w, (n_1m_1, ..., n_dm_d))$
\State $\mathcal{G}_1, \mathcal{G}_2, ..., \mathcal{G}_d = \Call{TTD}{W, (n_1m_1, ..., n_dm_d), r}$
\State \Return{$\mathcal{X}, \mathcal{G}_1, \mathcal{G}_2, ..., \mathcal{G}_d$} 

\Function{TTD}{$W, (n_1m_1, ..., n_dm_d), r$}\cite{oseledets2011tensor} 
    \State {Initialization}
    \State Temporary tensor: $C=W, r_0=1$;
    \For{each $k \in [1,d-1]$}
        \State $C=reshape(C,[r_{k-1}n_km_k, \frac{numel(C)}{r_{k-1}n_km_k} ])$
        \State Compute SVD: $C = USV+E$
        \State New core: $\mathcal{G}_k=reshape(U,[r_{k-1}, n_km_k, r_k])$
        \State $C=SV^T$
    \EndFor
    \State $\mathcal{G}_d = C$
    \State \Return{$\mathcal{G}_1, ... , \mathcal{G}_d$}
\EndFunction

\label{code:recentEnd}
\end{algorithmic}

\end{algorithm}

\section{TTD Inference of LLMs on GVSA}
\subsection{Operator Fusion of LLMs Inference}
For various operations in the LLMs inference process such as linear layers, batch normalization (BN) and residual connections (Res), operator fusion can be applied to merge two adjacent operators, reducing redundant data transfers, between off-chip and on-chip memory, thereby improving throughput. 

\begin{figure}[h]
\vspace{-0.4cm}
  \centering
  \includegraphics[width=0.9\linewidth]{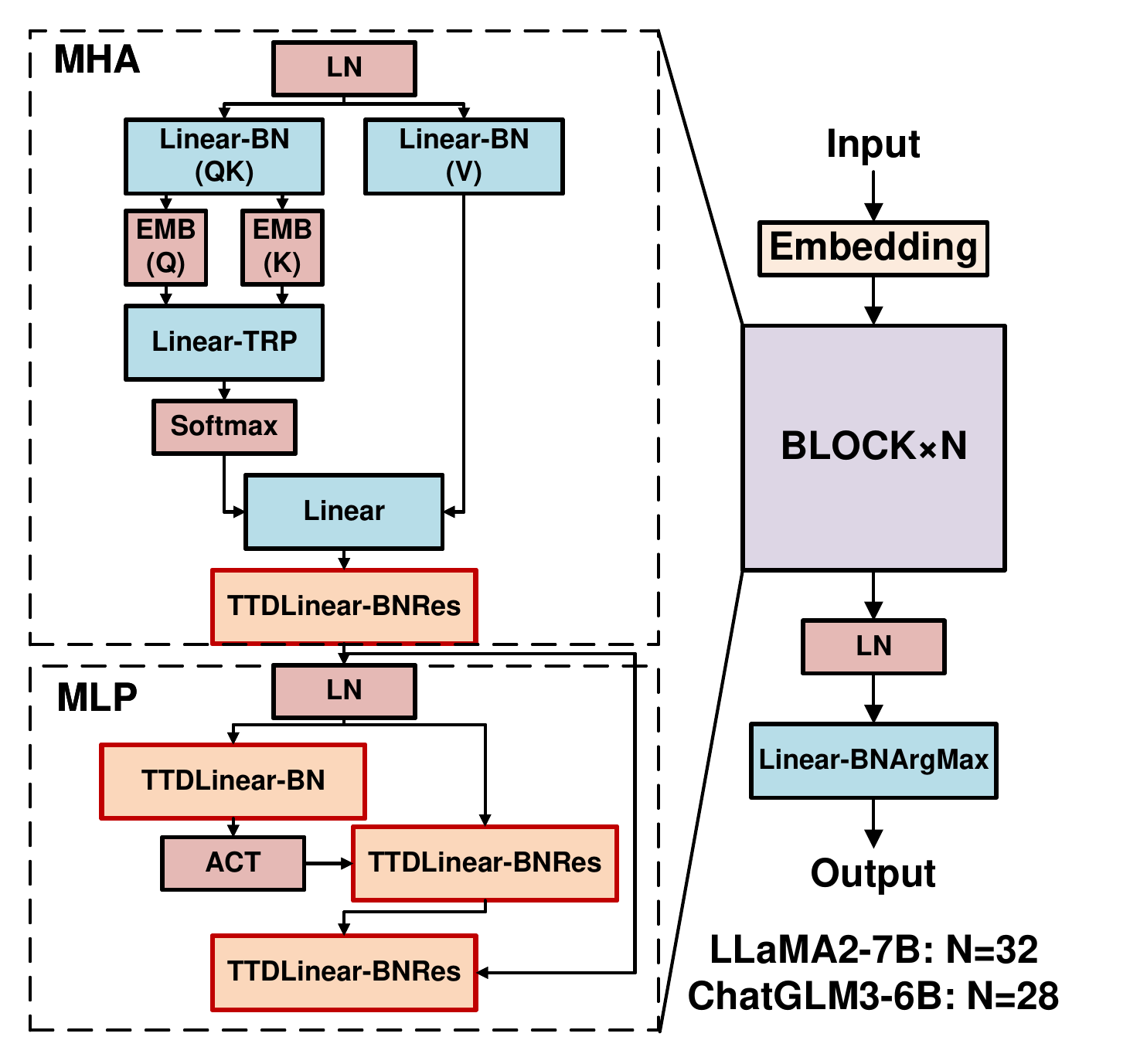}
  \caption{Operator execution order for inference of TTD compressed LLMs.}
  \label{fig_算子融合}
  \vspace{-0.4cm}
\end{figure}

Through operator fusion, the inference of TTD compressed LLMs is achieved by configuring three main operations in the accelerator: TTD linear operation, linear operation and nonlinear operation. The TTD linear operation focuses on TTD inference for the linear layer, integrating BN and Res operations. The linear operation handles the regular linear layer and related operations. The nonlinear operation is responsible for various nonlinear operations, such as layer normalization (LN), embedding (EMB), activation functions (ACT) and softmax. Fig.~\ref{fig_算子融合} shows the operators execution order for the inference of ChatGLM3-6B and LLaMA2-7B after TTD compression. 

The compiler generates serialized instructions based on the network architecture. These instructions configure operations by writing to registers, including the information of operations and the addresses of input and output data in HBM/DDR. The serialized instructions ensure the sequential chaining of operations and completing the inference of the entire network.

\subsection{GVSA Architecture}
\begin{figure}[h]
    \vspace{-0.4cm}
  \centering
  \includegraphics[width=0.95\linewidth]{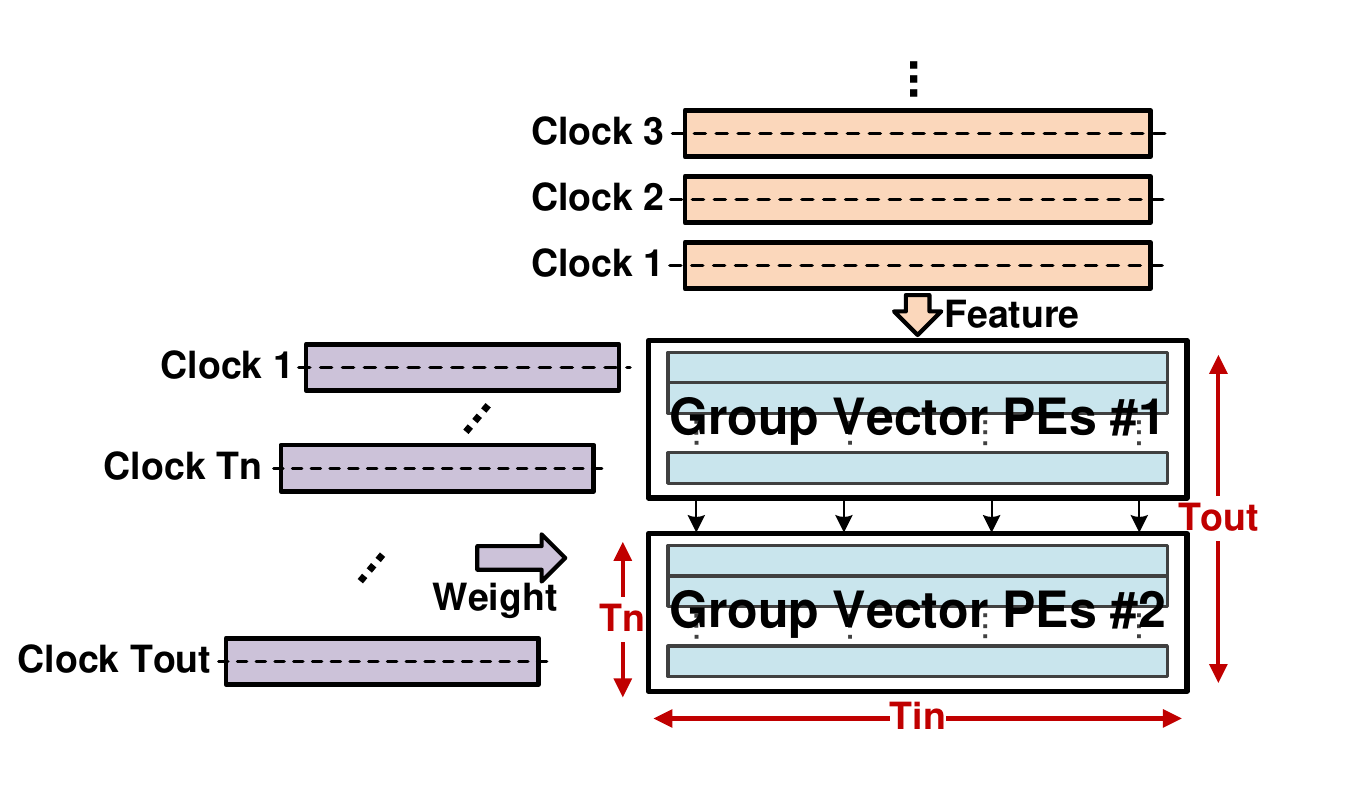}
  \caption{Architecture of GVSA when $T_n=T_{out}/2$.}
  \label{fig_GVSA}
  \vspace{-0.4cm}
\end{figure}

We utilize the GVSA architecture in\cite{huang2023integer} employing a multiple row-by-row systolic computing style to achieve the highly efficient matrix multiplication as shown in Fig.~\ref{fig_GVSA}. The array consists of $T_{out}$ PEs, each performing $T_{in}$ MAC operations. Parallel computation with $T_n$ PEs, executes MAC operations for $1 \times T_{in}$ feature with $T_n \times T_{in}$ weight. At each cycle, only one weight vector is loaded into this array. Meanwhile, all of others remain stationary within the PEs for $T_{out}$ cycles until they are replaced by a new one. Feature vectors are sequentially fed into the array and passed from one group of parallel vector PEs to the next.

\subsection{TTD Inference Data Mapping on GVSA}
For TTD inference, as shown in Equation~(\ref{eq_TTD}), if the computation is performed by multiplying cores by the input and then accumulating the input dimensions $i_k$ to obtain the final output, it will result in a lot of unnecessary memory access and computation. For example, in a 4-dimensional TTD inference, consider two output results $\mathcal{Y}(j_1,j_2,j_3,j_4)$ when their $j_1,j_2,j_3=0$ and one $j_4=1$, while the other $j_4=2$. If you need to calculate these two values, their cores $\mathcal{G}_1$, $\mathcal{G}_2$, $\mathcal{G}_3$ are all the same due to the same value of output dimension in $j_1,j_2,j_3$, and only cores $\mathcal{G}_4$ are different. Thus, the computation of the partial products and memory access for the first three cores is redundant. Therefore, a more efficient computation strategy becomes necessary.

\begin{figure}[h]
\vspace{-0.4cm}
  \centering
  \includegraphics[width=\linewidth]{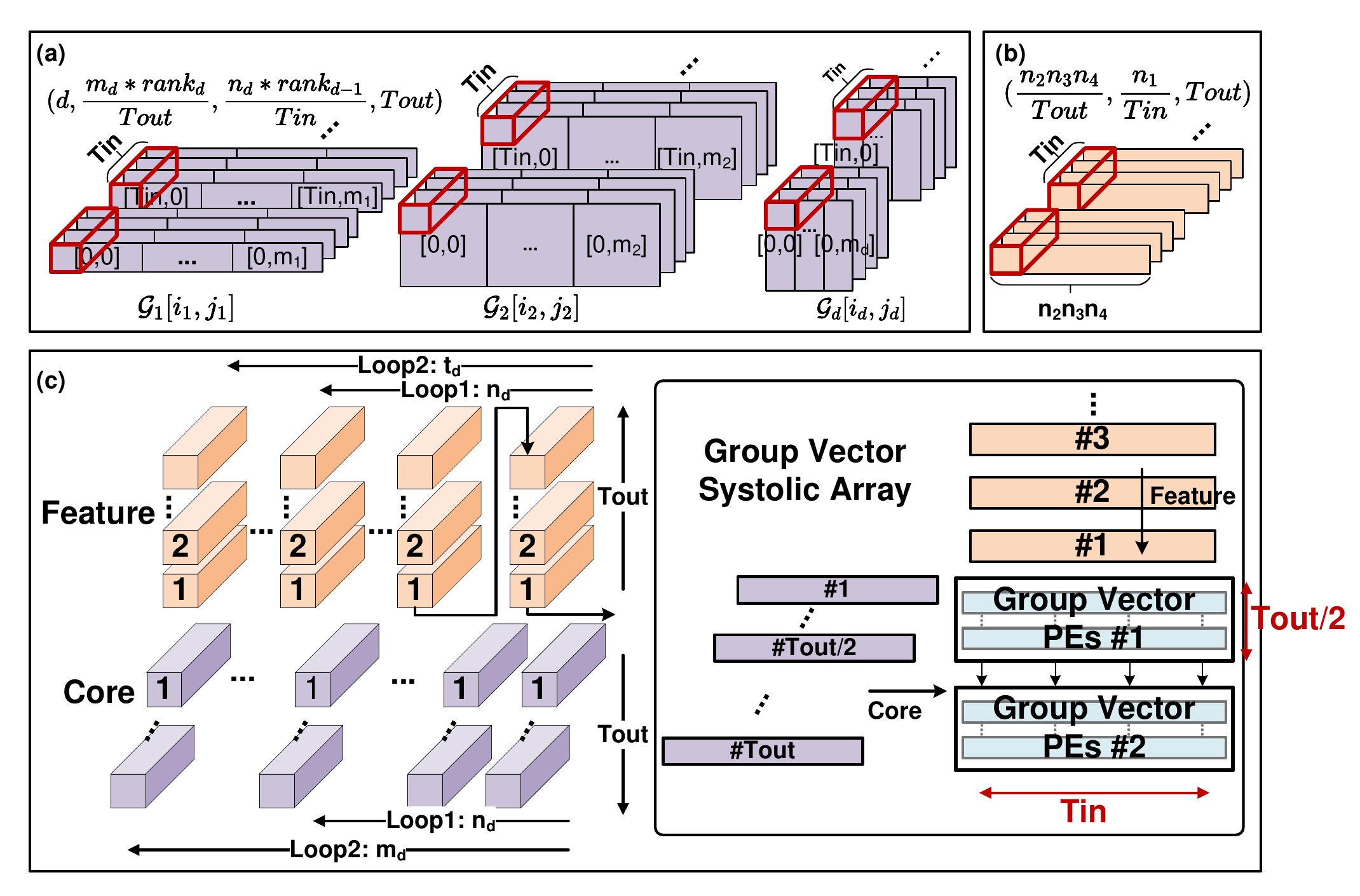}
  \caption{(a) Data mapping of cores; (b) Data mapping of input features; (c) Operation stages on GVSA for TTD inference.}
  \label{fig_TTD算子硬件结构}
\end{figure}

We first split the computation order, starting with the first input dimension $i_1$ and proceeding sequentially until summing over the final input dimension $i_d$. In this computation order, assuming we denote the partial computation results as $\mathcal{P}$, each layer's calculation involves multiplying the last partial result $\mathcal{P}_{k-1}$ with the current core $\mathcal{G}_k$ and then summing over the input dimension $i_k$, as shown in Equation~(\ref{eq_TTD_hw}).
\begin{equation}
\label{eq_TTD_hw}
\overline{\mathcal{P}_k}[t_{k-1},j_k]=\sum_{i_k}^{n_k}\mathcal{G}_k[i_k,j_k]\mathcal{P}_{k-1}[i_k,t_{k-1}]
\end{equation}
where $\mathcal{P}_0=\mathcal{X}$, and $\overline{\mathcal{P}_d}=\mathcal{Y}$. And $\overline{\mathcal{P}_k}[t_{k-1},j_k]$ is reordered to $\mathcal{P}_k[i_k, t_k]$. $t_k$ is the other dimension of stage $k$, such as, $t_0 = [i_2,i_3,i_4]$ (example with $d$=4) and $t_1 = [i_3,i_4,j_1]$. The following part will introduce the data mapping and data flow of the accelerator when executing TTD inference of linear layer.

\textbf{Data Preparation Stage.} The input $\mathcal{X}$ and weight cores $\mathcal{G}$ are input to the accelerator from the off-chip DDR and HBM through the AXI bus, and the DMA stores them in the buffer.

\textbf{Stage 1.} When the data is prepared for computation, it is processed in Stage 1 (S1), characterized by a multi-level loop calculation within GVSA. For the data flow of $\mathcal{X}$ into GVSA, where $i_1$  is the summation dimension and its parallelism is $T_{in}$, it will first complete the loop over $T_{out}$ on output dimension, and then complete the loop over the summation dimension which is $\lceil n_1 / T_{in} \rceil$ . Finally, it completes the loops over other dimensions which is $\lceil (n_2 \times n_3 \times n_4) / T_{out} \rceil$. For the TTD core $ \mathcal{G}_1 $, the process is similar: first, loop over $ T_{out} $, then complete the loop of the summation dimension, and finally loop over the output dimension which is $\lceil (m_1 \times \text{rank}) / T_{out} \rceil$. In S1, the control unit reads the input feature $\mathcal{X}$ and $ \mathcal{G}_1 $ core data from the input buffer. The output $ \overline{\mathcal{P}_1} $ is written into the ping-pong buffers. The shape of the cores and input feature and operation scheme with GVSA are shown in Fig.~\ref{fig_TTD算子硬件结构}.

\textbf{Stage 2.} Next, in Stage 2 (S2), the loops for the $ \mathcal{G}_2 $ core are similar to those in S1, with the exception that the summation dimension loop is $\lceil (n_2 \times \text{rank}) / T_{in} \rceil$. The input $ \mathcal{P}_1 $ undergoes reordering. In this process, since the summation dimension $ i_2 $ in S2 is positioned within the output dimension $ (i_2, i_3, i_4, j_1) $ of $\overline{\mathcal{P}_1}$ and it repeats in the time direction for each $ T_{out} $ cycle, it must be extracted and reconstituted into $ \mathcal{P}_1 $ with $ T_{in} $ parallelism to match the input format of GVSA. This reordering operation is embedded within the writing of $ \overline{\mathcal{P}_1} $ and the reading of $ \mathcal{P}_1 $.

The remaining stages are similar until all stages are completed. The results are input to BN and Res modules and then written to the external DDR by DMA.

\textbf{Reorder Between Stages.} We reorder $\overline{\mathcal{P}_1}$ to $\mathcal{P}_1$ by writing the the $T_{out}$-length vector-shaped output of S1 to the ping-pong buffers which has two dimensions: block dimension which is summation dimension in S2 and address dimension which is the time dimension in S2. In S2, the parts of the input $\mathcal{P}_1$ are read from the same address across different blocks of buffers. By such operations, the reordering process is hidden in the reading and writing of intermediate data between stages in the ping-pong buffers, avoiding the waiting time and eliminating the need for a reordering unit.

\input{LLM_benchmark}

\section{FPGA Implementation}
The architecture of the TTD linear operation in the accelerator as shown in the Fig.~\ref{fig_总体架构} comprises DDR/HBM Direct Memory Access (DMA), an input buffer, ping-pong buffers, a GVSA and a control unit. The DMA is responsible for transferring input features and cores of TTD from external DDR and HBM to the on-chip buffer, as well as moving the output data to the DDR. The ping-pong buffers store intermediate data in every stage of TTD inference. The control unit manages the reading and writing data from the buffers.

\begin{figure}[h]
  \centering
  \includegraphics[width=1.04\linewidth]{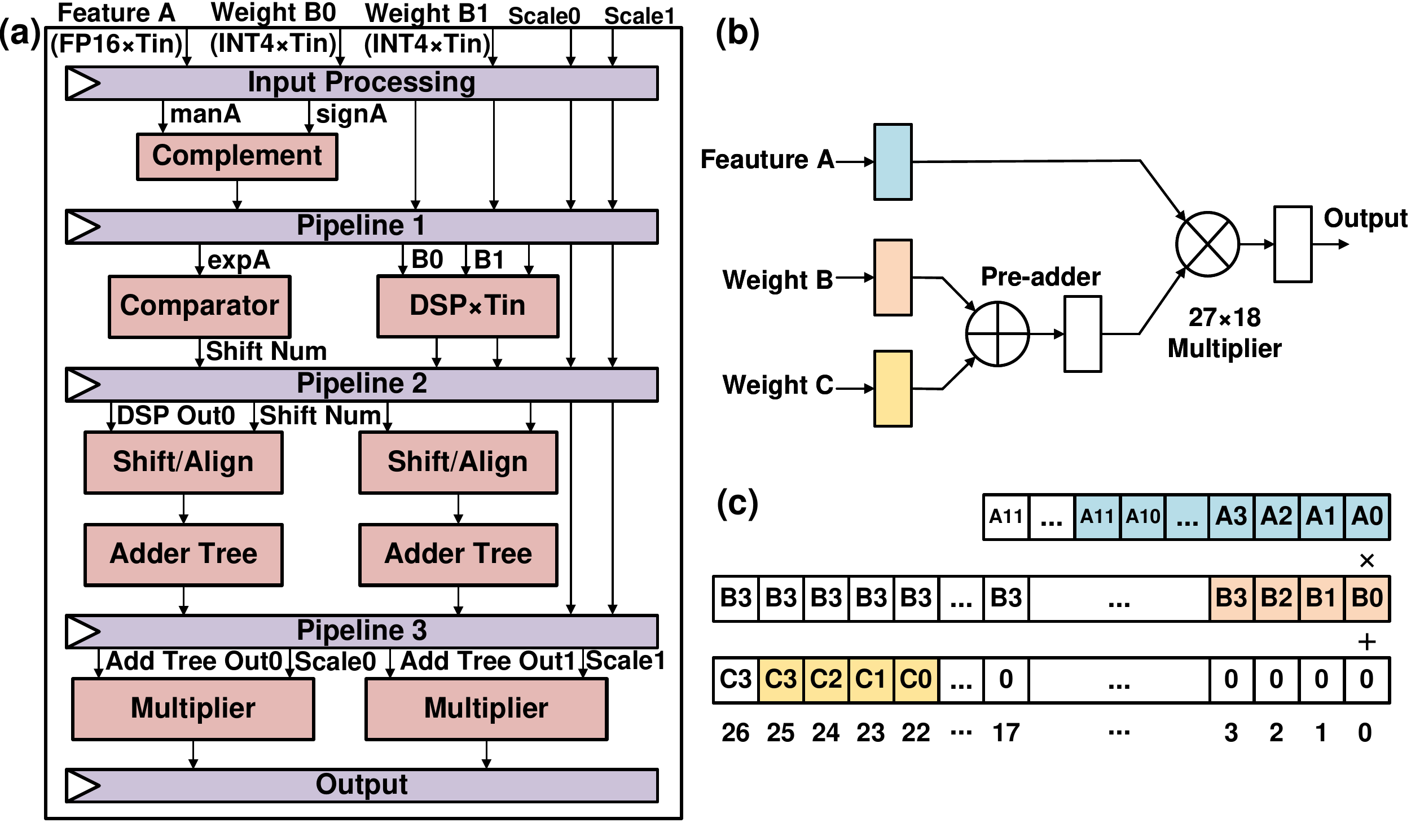}
  \caption{(a)DSP-shared GVSA vector PEs; (b)According design by DSP in Xilinx; (c) Data arrangement of DSP inputs.}
  \label{fig_PE}
  \vspace{-0.4cm}
\end{figure}

Because of the complex structure and various operations, efficiently utilizing limited resources is essential when deploying LLMs on FPGA. We propose DSP-shared floating-point parallel vector PEs to execute FP16$\times$INT4 operations, which not only adapts to the data flow of the GVSA but also optimizes the utilization of DSP resources. As shown in Fig.~\ref{fig_PE}(b), the multipliers in the DSPs on the Xilinx Ultrascale+ platform have input bit widths of 27 and 18. Previous works\cite{kalali2021near, huang2023multi, xue2022dual} have used DSPs for various integer calculations. By a similar approach,  we can achieve FP16$\times$INT4 computations through the arrangement of input data.

The FP16 format consists of a 1-bit sign, a 5-bit exponent, and a 10-bit mantissa. To perform FP16 and INT4 multiplication, we need to multiply the signed 4-bit fixed-point number by the 12-bit two's complement representation of $\{sign, hidden bit, mantissa\}$. As mentioned above, the adequate input bit width of the DSP allows us to simultaneously perform multiplication of a two's complement mantissa and two INT4 weights. The arrangement of DSP inputs is shown in Fig.~\ref{fig_PE}(c). In this way, the output of the DSP includes two 16-bit multiplication results. Through this DSP-shared strategy, we not only meet the data flow requirements of GVSA but also save DSP resources. By using multiple such PEs in parallel externally, we can achieve various levels of parallelism.

For the DSP-shared vector PEs, due to the multiple steps in floating-point calculations, we design it in a pipeline structure shown in Fig.~\ref{fig_PE}(a). One feature vector and two weight vectors are fed into the vector PEs. First, the parts of the FP16 format are separated, and the $\{sign, hidden bit, mantissa\}$ data passes a two's complement operation. Next, the 12-bit two's complement and the fixed-point weight data are fed into the DSPs. Simultaneously, the floating-point exponent bits are compared to determine the shift amount relative to the maximum exponent bit for subsequent adder trees. The DSP output is then shifted according to the determined shift amount and fed into the adder trees. In the final pipeline stage, the input scale is multiplied with the result of the accumulation.

\section{Experimental Results}

\subsection{Experiment Setup}
We used Vivado 2024.1 software to synthesize and implement our proposed accelerator on the AMD Alveo V80 FPGA platform with 12574K LUTs, 5148K FFs, 3741 Block RAMs (BRAMs) and 10848 DSPs. This FPGA integrates 32GB of HBM which is suitable for LLMs inference.

The benchmarks for this accelerator are TTD compressed ChatGLM3-6B and LLaMA2-7B. To ensure the accuracy of these networks, we choose specific blocks to perform TTD compression and INT4 quantization on the linear layers excluding $Attention(Q,K,V)$, while the remaining parts perform only INT4 quantization. These TTD compressed blocks are concatenated with the other quantized blocks and quantized output layers to construct the final deployed network. The configurations of the benchmarks are shown in the Table~\ref{tab_benchmark}. 

In deploying the inference process of TTD compressed LLMs, the operators of the accelerator are deployed and executed sequentially according to the operator chaining order in Fig.~\ref{fig_算子融合}, completing the accelerator system emulation for V80 on the benchmarks which is similar with the experiment setup of FlightLLM\cite{zeng2024flightllm}.


\input{V80_FPGA_Resource}

\subsection{Network Performance Evaluation}
The CRs and performance of TTD compressed LLMs are shown in Table~\ref{tab_benchmark}. The number of TTD compressed blocks of ChatGLM3-6B and LLaMA2-7B is 15 and 19, respectively. The TTD compression for ChatGLM3-6B and LLaMA2-7B achieves 481.88$\sim$1446.44$\times$ CRs for linear layers, 10.72$\times$ and 4.01$\times$ CRs for single block, 1.94$\times$ and 1.60$\times$ CRs for whole network, respectively. Compared to the original model, the average score on C-EVal tests\cite{huang2023ceval} of ChatGLM3-6B after TTD compression and quantization shows a decrease of 4.21. The perplexity (PPL) on C4 dataset\cite{2019t5} of LLaMA2-7B shows an increase of 2.62.

\subsection{Hardware Performance Evaluation}
\begin{figure}[t!]
  \centering
  \includegraphics[width=\linewidth]{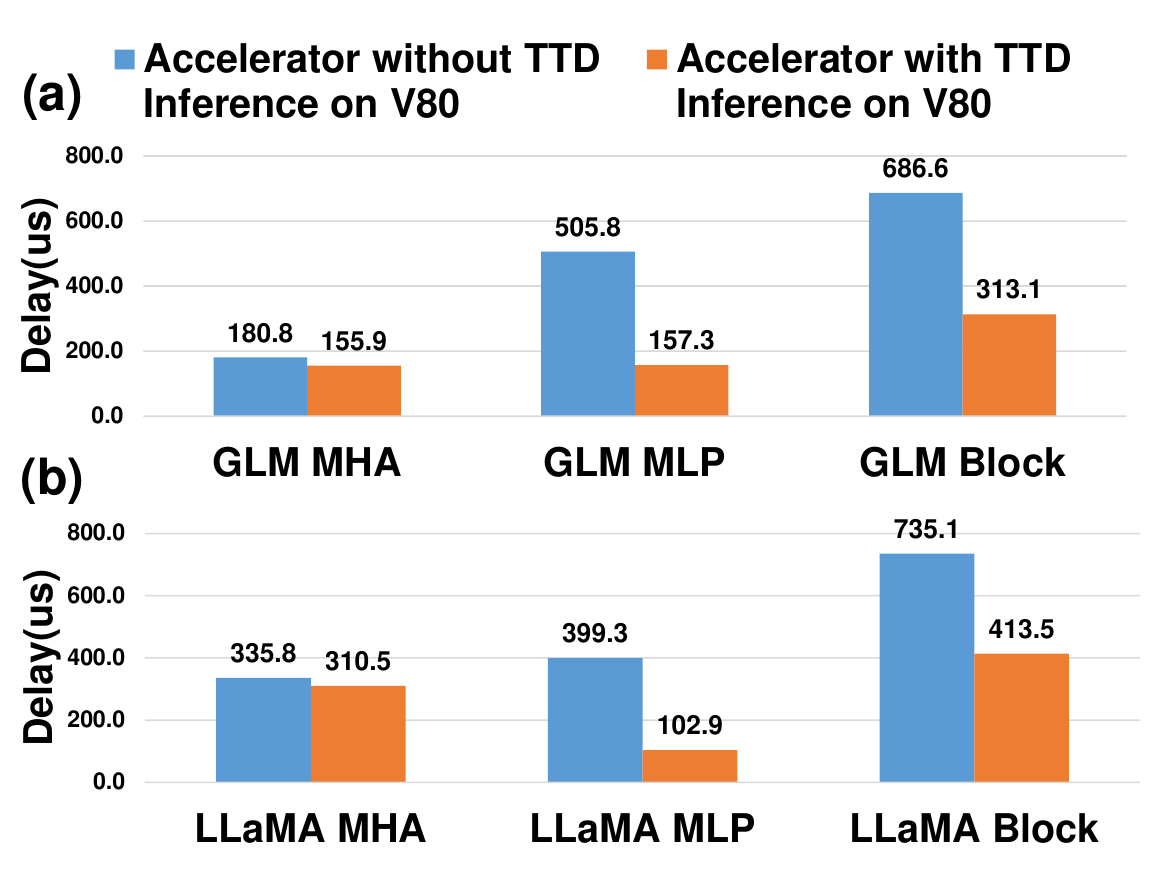}
  \caption{Performance comparison with accelerator without TTD inference on (a) ChatGLM3-6B and (b) LLaMA2-7B.}
  \label{fig_对比图1}
\end{figure}

For GVSA in the TTD linear operation, $T_{in}$, $T_{out}$ and $T_n$ are set to 128, 32 and 16, respectively. The frequency of system operation of this accelerator is set at 125MHz. The resource utilization is shown in Table~\ref{tab_fpga_res}. The deployment of the ChatGLM3-6B and LLaMA2-7B on the accelerator is shown in Table~\ref{tab_glm_delay} and Table~\ref{tab_llama_dalay}, respectively. For this accelerator, we map the inference process of single block in ChatGLM3-6B and LLaMA2-7B into a 14 steps of operators, and output layers mapped into 2 steps. We also use a same accelerator but without TTD inference as a baseline. As shown in Fig.~\ref{fig_对比图1}, the accelerator achieves 3.22$\times$ and 3.88$\times$ speedup in the MLP inference, 2.19$\times$ and 1.78$\times$ speedup in the single block inference and 1.45$\times$ and 1.57$\times$ reduction in the first token delay of ChatGLM3-6B and LLaMA2-7B, respectively.
 
\input{GLM_delay_V80}

\input{LLaMA_delay_V80}

\begin{figure}[t!]
  \centering
  \includegraphics[width=1.05\linewidth]{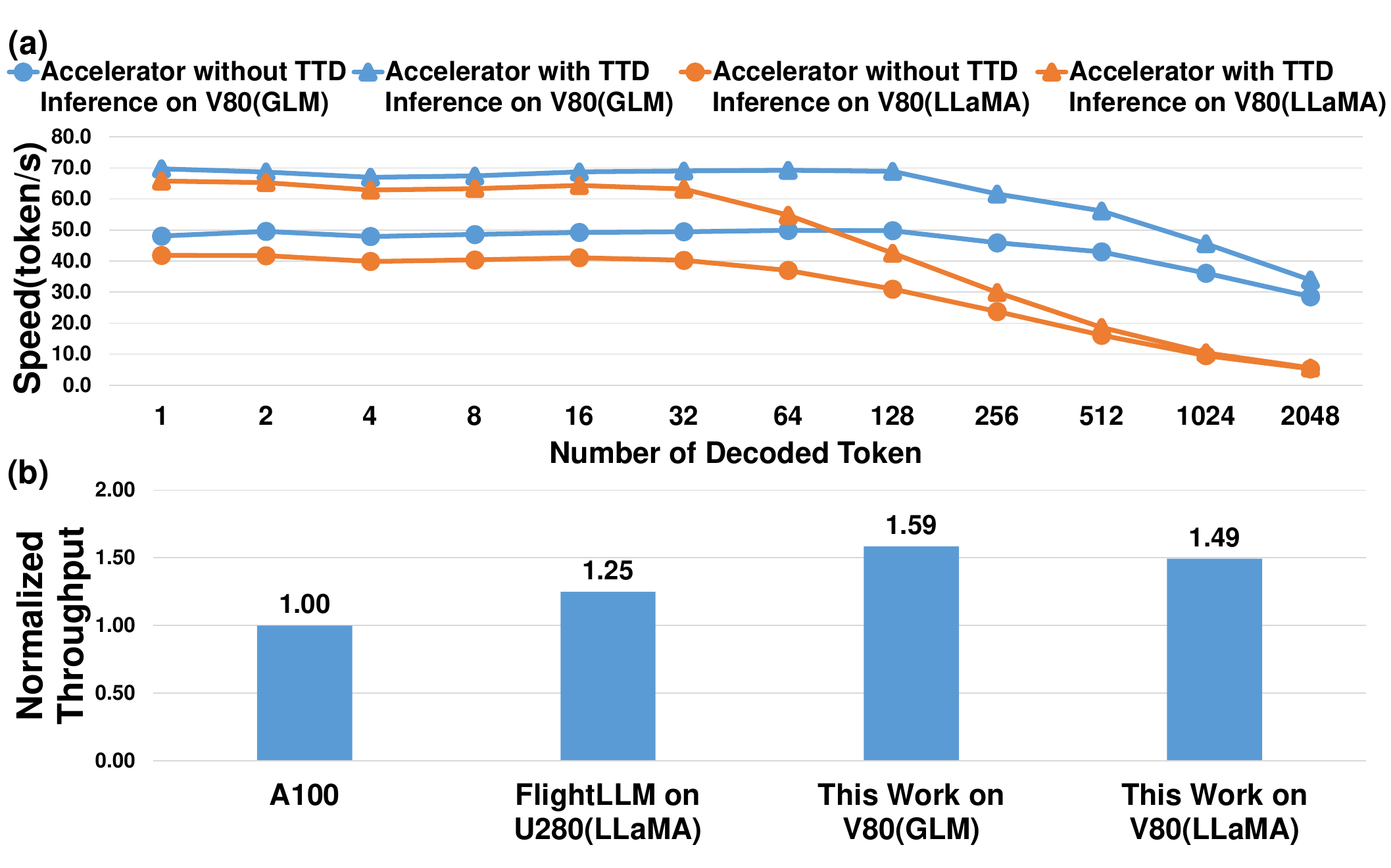}
  \caption{(a) Speed comparison with accelerator without TTD inference at different token number in decoding stage of ChatGLM3-6B and LLaMA2-7B. (b) Normalized throughput comparison with different hardware platform. }
  \label{fig_对比图2}
  \vspace{-0.4cm}
\end{figure}

Fig.~\ref{fig_对比图2}(a) shows the speed of the accelerator when executing ChatGLM3-6B and LLaMA2-7B for different number of decoded tokens. It can be seen that the KV cache strategy mitigates the decrease in speed when generating longer statements and the accelerator with TTD inference always has better performance than the baseline. Fig.~\ref{fig_对比图2}(b) shows the normalized throughput of the accelerator compared to other hardware platforms. The accelerator with TTD inference achieves 59\% and 49\% higher throughput compared to the A100 GPU. FlightLLM is a current advanced LLM accelerator that achieves good throughput through sparsity and mixed precision. By applying TTD compression to LLMs, optimizing the data flow of TTD inference and using highly efficient GVSA, our accelerator achieves a 27\% and 19\% higher throughput compared to FlightLLM on U280.

\section{Conclusion}
In this paper, we implement TTD compressed LLMs and deploy their inference on FPGA. The TTD inference is mapped onto a highly efficient GVSA with DSP-shared parallel vector PEs. Experimental results show that the proposed accelerator achieves a 3.22$\times$ and 3.88$\times$ speedup in the MLP inference, 2.19$\times$ and 1.78$\times$ speedup in the single block inference and 1.45$\times$ and 1.57$\times$ reduction in first token delay on ChatGLM3-6B and LLaMA2-7B, respectively.

\clearpage

\bibliographystyle{IEEEtran}
\bibliography{IEEEabrv,ref}

\end{document}

%% file: LLM_benchmark.tex
\begin{table*}
\caption{Information of evaluated benchmarks for the accelerator}
\label{tab_benchmark}
\large
\centering
\resizebox{\textwidth}{!}{

\begin{tabular}{|c|cccccc|c|c|c|c|c|c|}
\hline
\textbf{Benchmark}                                                           & \multicolumn{2}{c}{\textbf{Operations}} & \textbf{Size} & \textbf{n}      & \textbf{m}      & \textbf{Rank} & \textbf{Precision}                                                             & \textbf{\begin{tabular}[c]{@{}c@{}}CR for\\ Linear Layer\end{tabular}} & \textbf{\begin{tabular}[c]{@{}c@{}}CR for\\ Single Block\end{tabular}} & \textbf{\begin{tabular}[c]{@{}c@{}}CR for\\ Whole Network\end{tabular}} & \textbf{\begin{tabular}[c]{@{}c@{}}Score\\ Decrease\end{tabular}} & \textbf{\begin{tabular}[c]{@{}c@{}}PPL\\ Increase\end{tabular}} \\ \hline
\multirow{7}{*}{\begin{tabular}[c]{@{}c@{}}ChatGLM3-6B\\ Block\end{tabular}} & \multirow{4}{*}{MHA}      & LinearQ     & 4096$\times$4096     & -               & -               & -             & \multirow{14}{*}{\begin{tabular}[c]{@{}c@{}}Wt: INT4\\ Act: FP16\end{tabular}} & 1                                                                      & \multirow{7}{*}{10.72}                                                 & \multirow{7}{*}{1.94}                                                   & \multirow{7}{*}{4.21}                                          & \multirow{7}{*}{-}                                                  \\
                                                                             &                           & LinearK     & 4096$\times$256      & -               & -               & -             &                                                                                & 1                                                                      &                                                                        &                                                                         &                                                                  &                                                                     \\
                                                                             &                           & LinearV     & 4096$\times$256      & -               & -               & -             &                                                                                & 1                                                                      &                                                                        &                                                                         &                                                                  &                                                                     \\
                                                                             &                           & LinearO     & 4096$\times$4096     & {[}16,8,8,4{]}  & {[}4,8,8,16{]}  & 16            &                                                                                & 481.88                                                                 &                                                                        &                                                                         &                                                                  &                                                                     \\ \cline{2-7} \cline{9-9}
                                                                             & \multirow{3}{*}{MLP}      & Linear1     & 4096$\times$13696    & {[}8,8,8,8{]}   & {[}4,4,8,107{]} & 16            &                                                                                & 1446.44                                                                &                                                                        &                                                                         &                                                                  &                                                                     \\
                                                                             &                           & Linear2     & 4096$\times$13696    & {[}8,8,8,8{]}   & {[}4,4,8,107{]} & 16            &                                                                                & 1446.44                                                                &                                                                        &                                                                         &                                                                  &                                                                     \\
                                                                             &                           & Linear3     & 13696$\times$4096    & {[}107,8,4,4{]} & {[}8,8,8,8{]}   & 16            &                                                                                & 1446.44                                                                &                                                                        &                                                                         &                                                                  &                                                                     \\ \cline{1-7} \cline{9-13} 
\multirow{7}{*}{\begin{tabular}[c]{@{}c@{}}LLaMA2-7B\\ Block\end{tabular}}   & \multirow{4}{*}{MHA}      & LinearQ     & 4096$\times$4096     & -               & -               & -             &                                                                                & 1                                                                      & \multirow{7}{*}{4.01}                                                  & \multirow{7}{*}{1.60}                                                   & \multirow{7}{*}{-}                                               & \multirow{7}{*}{2.62}                                                \\
                                                                             &                           & LinearK     & 4096$\times$4096     & -               & -               & -             &                                                                                & 1                                                                      &                                                                        &                                                                         &                                                                  &                                                                     \\
                                                                             &                           & LinearV     & 4096$\times$4096     & -               & -               & -             &                                                                                & 1                                                                      &                                                                        &                                                                         &                                                                  &                                                                     \\
                                                                             &                           & LinearO     & 4096$\times$4096     & {[}16,8,8,4{]}  & {[}4,8,8,16{]}  & 16            &                                                                                & 481.88                                                                 &                                                                        &                                                                         &                                                                  &                                                                     \\ \cline{2-7} \cline{9-9}
                                                                             & \multirow{3}{*}{MLP}      & Linear1     & 4096$\times$11008    & {[}16,8,8,4{]}  & {[}4,4,16,43{]} & 16            &                                                                                & 1233.82                                                                &                                                                        &                                                                         &                                                                  &                                                                     \\
                                                                             &                           & Linear2     & 4096$\times$11008    & {[}16,8,8,4{]}  & {[}4,4,16,43{]} & 16            &                                                                                & 1233.82                                                                &                                                                        &                                                                         &                                                                  &                                                                     \\
                                                                             &                           & Linear3     & 11008$\times$4096    & {[}43,16,4,4{]} & {[}4,8,8,16{]}  & 16            &                                                                                & 1007.89                                                                &                                                                        &                                                                         &                                                                  &                                                                     \\ \hline
\end{tabular}

}
\vspace{-0.2cm}
\end{table*}

%% file: V80_FPGA_Resource.tex



\begin{table}[h]
\vspace{-0.2cm}
\caption{Resource utilization of the accelerator on V80}
\label{tab_fpga_res}
\centering
\resizebox{\linewidth}{!}{

\begin{tabular}{|c|c|c|c|c|}
\hline
\textbf{Resource} & \textbf{Available} & \textbf{\begin{tabular}[c]{@{}c@{}}TTD Linear\\ Operation\end{tabular}} & \textbf{\begin{tabular}[c]{@{}c@{}}Other\\ Operations\end{tabular}} & \textbf{Total} \\ \hline
LUT               & 2574K              & 403K(15.68\%)                                                           & 903K                                                                & 1307K(50.79\%) \\ \hline
FF                & 5148K              & 349K(6.78\%)                                                            & 569K                                                                & 919K(17.85\%)  \\ \hline
BRAM              & 3741               & 643(17.19\%)                                                            & 447                                                                 & 1090(29.14\%)  \\ \hline
DSP               & 10848              & 2159(19.90\%)                                                           & 7286                                                                & 9445(87.07\%)  \\ \hline
\end{tabular}

}
\vspace{-0.4cm}
\end{table}

%% file: GLM_delay_V80.tex
\begin{table}[t]
\caption{TTD compressed ChatGLM3-6B on V80}
\label{tab_glm_delay}
\centering

\begin{tabular}{|c|ccc|}
\hline
\textbf{ChatGLM3-6B}          & \multicolumn{2}{c|}{\textbf{Operations}}                                                              & \textbf{Delay(us)}      \\ \hline
\multirow{14}{*}{Block}       & \multicolumn{1}{c|}{\multirow{9}{*}{MHA}}       & \multicolumn{1}{c|}{LN}                             & 11.39                   \\ \cline{3-4} 
                              & \multicolumn{1}{c|}{}                           & \multicolumn{1}{c|}{Linear-BN(QK)}                  & 51.03                   \\ \cline{3-4} 
                              & \multicolumn{1}{c|}{}                           & \multicolumn{1}{c|}{EMB(Q)}                         & 6.54                    \\ \cline{3-4} 
                              & \multicolumn{1}{c|}{}                           & \multicolumn{1}{c|}{EMB(K)}                         & 6.80                    \\ \cline{3-4} 
                              & \multicolumn{1}{c|}{}                           & \multicolumn{1}{c|}{Linear-TRP}                     & 8.24                    \\ \cline{3-4} 
                              & \multicolumn{1}{c|}{}                           & \multicolumn{1}{c|}{Softmax}                        & 26.08                   \\ \cline{3-4} 
                              & \multicolumn{1}{c|}{}                           & \multicolumn{1}{c|}{Linear-BN(V)}                   & 7.47                    \\ \cline{3-4} 
                              & \multicolumn{1}{c|}{}                           & \multicolumn{1}{c|}{Linear}                         & 8.99                    \\ \cline{3-4} 
                              & \multicolumn{1}{c|}{}                           & \multicolumn{1}{c|}{\textbf{TTDLinear-BNRes}}       & \textbf{29.32}          \\ \cline{2-4} 
                              & \multicolumn{1}{c|}{\multirow{5}{*}{MLP}}       & \multicolumn{1}{c|}{LN}                             & 11.63                   \\ \cline{3-4} 
                              & \multicolumn{1}{c|}{}                           & \multicolumn{1}{c|}{\textbf{TTDLinear-BN}}          & \textbf{43.04}          \\ \cline{3-4} 
                              & \multicolumn{1}{c|}{}                           & \multicolumn{1}{c|}{ACT}                            & 21.87                   \\ \cline{3-4} 
                              & \multicolumn{1}{c|}{}                           & \multicolumn{1}{c|}{\textbf{TTDLinear-BNRes}}       & \textbf{43.49}          \\ \cline{3-4} 
                              & \multicolumn{1}{c|}{}                           & \multicolumn{1}{c|}{\textbf{TTDLinear-BNRes}}       & \textbf{37.22}          \\ \hline
\multirow{2}{*}{Output Layer} & \multicolumn{2}{c|}{LN}                                                                               & 13.78                   \\ \cline{2-4} 
                              & \multicolumn{2}{c|}{Linear-BNArgmax}                                                                  & 701.18                  \\ \hline
Summary                       & \multicolumn{3}{c|}{\begin{tabular}[c]{@{}c@{}}First token delay: 14.34ms\\ Peak speed: 69.7token/s(1.45$\times$)\end{tabular}} \\ \hline
\end{tabular}
\vspace{-0.4cm}
\end{table}

%% file: LLaMA_delay_V80.tex
\begin{table}[t]
\caption{TTD compressed LLaMA2-7B on V80}
\label{tab_llama_dalay}
\centering

\begin{tabular}{|c|ccc|}
\hline
\textbf{LLaMA2-7B}            & \multicolumn{2}{c|}{\textbf{Operations}}                                                         & \textbf{Delay(us)}    \\ \hline
\multirow{14}{*}{Block}       & \multicolumn{1}{c|}{\multirow{9}{*}{MHA}}     & \multicolumn{1}{c|}{LN}                          & 12.57                 \\ \cline{3-4} 
                              & \multicolumn{1}{c|}{}                         & \multicolumn{1}{c|}{Linear-BN(QK)}               & 91.23                 \\ \cline{3-4} 
                              & \multicolumn{1}{c|}{}                         & \multicolumn{1}{c|}{EMB(Q)}                      & 4.82                  \\ \cline{3-4} 
                              & \multicolumn{1}{c|}{}                         & \multicolumn{1}{c|}{EMB(K)}                      & 6.80                  \\ \cline{3-4} 
                              & \multicolumn{1}{c|}{}                         & \multicolumn{1}{c|}{Linear-TRP}                  & 47.35                 \\ \cline{3-4} 
                              & \multicolumn{1}{c|}{}                         & \multicolumn{1}{c|}{Softmax}                     & 22.35                 \\ \cline{3-4} 
                              & \multicolumn{1}{c|}{}                         & \multicolumn{1}{c|}{Linear-BN(V)}                & 51.94                 \\ \cline{3-4} 
                              & \multicolumn{1}{c|}{}                         & \multicolumn{1}{c|}{Linear}                      & 44.13                 \\ \cline{3-4} 
                              & \multicolumn{1}{c|}{}                         & \multicolumn{1}{c|}{\textbf{TTDLinear-BNRes}}    & \textbf{29.34}        \\ \cline{2-4} 
                              & \multicolumn{1}{c|}{\multirow{5}{*}{MLP}}     & \multicolumn{1}{c|}{LN}                          & 11.00                 \\ \cline{3-4} 
                              & \multicolumn{1}{c|}{}                         & \multicolumn{1}{c|}{\textbf{TTDLinear-BN}}       & \textbf{27.03}        \\ \cline{3-4} 
                              & \multicolumn{1}{c|}{}                         & \multicolumn{1}{c|}{ACT}                         & 12.43                 \\ \cline{3-4} 
                              & \multicolumn{1}{c|}{}                         & \multicolumn{1}{c|}{\textbf{TTDLinear-BNRes}}    & \textbf{27.74}        \\ \cline{3-4} 
                              & \multicolumn{1}{c|}{}                         & \multicolumn{1}{c|}{\textbf{TTDLinear-BNRes}}    & \textbf{24.73}        \\ \hline
\multirow{2}{*}{Output Layer} & \multicolumn{2}{c|}{LN}                                                                          & 12.28                 \\ \cline{2-4} 
                              & \multicolumn{2}{c|}{Linear-BNArgmax}                                                             & 349.16                \\ \hline
Summary                       & \multicolumn{3}{c|}{\begin{tabular}[c]{@{}c@{}}First token delay: 15.20ms\\ Peak speed: 65.8token/s(1.57$\times$)\end{tabular}} \\ \hline
\end{tabular}
\vspace{-0.4cm}
\end{table}